\documentclass[reqno]{amsart}
\usepackage{times}
\newtheorem{theorem}{Theorem}[section]
\newtheorem{proposition}[theorem]{Proposition}
\newtheorem{lemma}[theorem]{Lemma}
\newtheorem{corollary}[theorem]{Corollary}
\theoremstyle{definition}
\newtheorem{remark}[theorem]{Remark}

\def\esssup{\mathop{\hbox{ess$\,$sup}}}

\def\supp{\mathop{\hbox{supp}}}
\newcommand{\Hmm}[1]{\leavevmode{\marginpar{\tiny%
$\hbox to 0mm{\hspace*{-0.5mm}$\leftarrow$\hss}%
\vcenter{\vrule depth 0.1mm height 0.1mm width \the\marginparwidth}%
\hbox to 0mm{\hss$\rightarrow$\hspace*{-0.5mm}}$\\\relax\raggedright #1}}}
\makeatletter
\newcounter{numcount}
\newcommand{\labelnummer}{\mbox{(\roman{numcount})}}%
    {\let\curlabelspeicher\@currentlabel%
     \begin{list}{\labelnummer}{\usecounter{numcount}\leftmargin0em%
                  \topsep1ex\partopsep2ex\parsep0pt\itemsep0.5ex%\@plus1\p@%
                  \labelwidth2.5em\itemindent3.5em\labelsep1em}%
     \let\saveitem\item%
     \def\item{\saveitem%
               \def\@currentlabel{\curlabelspeicher\labelnummer}%
               \let\label\bemlabel}}%
   {\end{list}}%

\def\itemref#1{\expandafter\@setref\csname r@#1item\endcsname\@firstoftwo{#1}}%

\def\bemlabel#1{\@bsphack%
  \protected@write\@auxout{}%
         {\string\newlabel{#1}{{\@currentlabel}{\thepage}}}%
  \ifmmode\else%
  \protected@write\@auxout{}%
         {\string\newlabel{#1item}{{\labelnummer}{\thepage}}}%
  \fi%
  \@esphack}%

\makeatother
\begin{document}
\title[Magnetic Lifshits tails]{Classical magnetic Lifshits tails in three space dimensions:
impurity potentials with slow anisotropic decay}

\author{Dirk Hundertmark}
\address{Institut Mittag-Leffler, Aurav\"agen 17, S-182 60 Djursholm, Sweden.
                On leave from: Department of Mathematics,
                University of Illinois at Urbana-Champaign,
                1409 W.~Green Street,
                Urbana, IL 61801.}%
\email{dirk@math.uiuc.edu}%
\author{Werner Kirsch}
\address{Institut f\"ur Mathematik,
          Ruhr-Universit\"at Bochum, D--44780 Bochum, Germany}%
\email{werner.kirsch@mathphys.ruhr-uni-bochum.de}
\author{Simone Warzel}
\address{Institut f\"ur Theoretische Physik,
                       Universit\"at
                       Erlangen-N\"urnberg, Staudtstra{\ss}e 7, D--91058
                       Erlangen, Germany}%
\email{simone.warzel@physik.uni-erlangen.de}

\begin{abstract}
  We determine the leading low-energy fall-off of the integrated density of states of
  a magnetic Schr\"odinger operator with repulsive Poissonian random potential in case
  its single impurity potential has a slow anisotropic decay at infinity.
  This so-called magnetic Lifshits tail is shown to coincide with the
  one of the corresponding classical integrated density of states.
\end{abstract}
\maketitle

\section{Introduction}
Random one-particle Schr\"odinger operators with (constant) magnetic fields have been attracting
considerable attention in the physics
as well as mathematics community. Physically speaking,
each of these operators models a spinless quantum particle which moves in the Euclidean configuration space $ \mathbb{R}^3 $
subject to a random potential $ V_\omega : \mathbb{R}^3 \to \mathbb{R} $ and a
constant magnetic field of strength $ B > 0 $.
In physical units where Planck's constant (divided by $2 \pi $) as well as the mass and the charge of the particle are all equal one,
the corresponding Schr\"odinger operator is informally given by the differential expression
\begin{equation}\label{def:H}
  H(V_\omega) := \frac{1}{2} \sum_{j=1}^3 \left( i \frac{\partial}{\partial x_j} + A_j \right)^2 + V_\omega
\end{equation}
which acts on the Hilbert space $ \mathrm{L}^2(\mathbb{R}^3) $ of complex-valued square-integrable functions on~$ \mathbb{R}^3 $.
Without loss of generality
one may choose co-ordinates $ x = (x_1, x_2, x_3) \in \mathbb{R}^3 $
such that the constant magnetic field is parallel to the $ x_3 $-axis. On account of gauge equivalence the
vector potential $ A:  \mathbb{R}^3 \to \mathbb{R}^3  $ in (\ref{def:H})
may therefore be fixed in the symmetric gauge by $  A(x) := \frac{B}{2} (-x_2 , x_1, 0) $.
In this paper the random potential $ V_\omega $ is supposed to be a repulsive Poissonian one for which
\begin{equation}\label{def:V}
  V_\omega(x) := \sum_{j} \, U\big(x - p_{\omega}(j) \big) \geq 0.
\end{equation}
Here for given realization $ \omega \in \Omega $ of the randomness,
the point $ p_{\omega}(j) \in \mathbb{R}^3 $ stands for the position of the $ j $th impurity
repelling the particle by a positive potential $ U \geq 0 $ which neither depends on $ \omega $
nor on $ j $. The impurities are distributed at a mean concentration $ \varrho > 0 $
according to Poisson's law such that the probability of simultaneously finding
$ M_1, M_2, \dots , M_K $ impurity points in respective pairwise disjoint
subsets $ \Lambda_1, \Lambda_2, \dots , \Lambda_3 \subset \mathbb{R}^3 $  is given by the product
$\prod_{k=1}^K e^{-\varrho \left| \Lambda_k \right|}
\left(\varrho \left| \Lambda_k \right| \right)^{M_k}/M_k! $, where
$ \left|\Lambda_k\right|:=\int_{\Lambda_k}\! d x$ is the volume of $\Lambda_k$.

The object of interest in this paper is the integrated density of states $ N $ of the Schr\"odinger operator
(\ref{def:H}) with Poissonian random potential (\ref{def:V}).
Informally, $ N (E) $ is just the number of energy levels per volume below a given energy $ E \in \mathbb{R} $.
See (\ref{def:N}) below and \cite{Pas73,Kir89,PaFi92} for an exact definition and general properties (in the case $B=0$).

Under some rather weak additional assumptions on $ U $ (see e.g.\ (\ref{Ass1}) below)
the almost-sure spectrum
of $ H(V_\omega) $ as well
as the set of growth points of its integrated density $ N $ are known to coincide
with the half-line $ [B/2 , \infty [ $.
We will investigate the behaviour of $ N $ near the bottom
of this half-line. More precisely, we will determine the
so-called \emph{magnetic Lifshits tail} of $ N $, that is,
the leading low-energy fall-off of $ N(E) $ as $ E \downarrow B/2 $.

Magnetic Lifshits tails have been investigated so far mainly for two space dimensions
in which two qualitatively different regimes were found \cite{BHKL95,Erd98,HLW99,HLW00,Erd01}.
Here for long-range $ U $ the Lifshits tails solely depend on the details of the decay of $ U $
and coincide with the low-energy fall-off of the corresponding classical integrated density of states.
For short-range $ U $, the tails are insensitive to the details of the decay of $ U $, but sensitive
on the magnetic field strength and have therefore a quantum character.
The borderline decay between such classical and quantum Lifshits tailing
in two space dimensions
has been shown to be Gaussian decay of $ U $.
This stands in contrast to the non-magnetic case in which algebraic decay
$ \lim_{| x | \to \infty } | x |^{\alpha} \, U(x) = g > 0  $ with exponent $\alpha= d + 2 $
discriminates between classical and quantum Lifshits tails in $ d $ space dimensions \cite{DonVar75,Pas77,Nak77}.
For a recent summary, see \cite[Sec.~4.1]{LMW02}.

First rigorous results on magnetic Lifshits tails in three space dimensions with rapidly decaying $ U $ are
available in \cite{War01}. The findings there especially reveal a regime of Lifshits tails with mixed classical
and quantum character for an intermediate decay of $ U $.
They will be published in an accompanying paper.
The purpose of this paper is to detect the regime of purely classical tails which we will prove to occur for
slow (anisotropic) decay of $ U $.
In particular, we will show that classical Lifshits tails exist \emph{at all} in three space dimensions.
This might be surprising from a naive point of view, since one may be tempted to argue that
the motion perpendicular to the magnetic field is confined and the particle can move freely
only parallel to the field lines such that the
effective dimension of the problem should be $ d = 1 $.
The regime of classical Lifshits tailing might therefore be
expected for algebraically decaying $ U $ with exponent $\alpha < 3 \, (= 1 + 2 )$. In the latter case, however,
the Poissonian random potential~(\ref{def:V}) is not well-defined in three space dimensions.
Thus, from this point of view,
one is lead to the wrong conclusion that classical Lifshits tails do not exist in three dimensions.

\section{Assumptions, definitions and results}
%
%\subsection{Assumptions and definitions}
%
Throughout this paper, we will consider non-negative impurity potentials $ U: \mathbb{R}^3 \to [0, \infty[ $
which are integrable as well as
square-integrable with respect to the three-dimensional Lebesgue measure
\begin{equation}\label{Ass1}
    U \geq 0, \qquad U \in {\rm L}^1\big(\mathbb{R}^3\big) \cap {\rm L}^2\big(\mathbb{R}^3\big).
\end{equation}
In particular, this ensures that the Poissonian random potential (\ref{def:V}) is
a positive, measurable, ergodic random field on some complete probability space
$ ( \Omega, \mathcal{A} , \mathbb{P} ) $.  Moreover, the operator $ H(V_\omega) $
is $ \mathbb{P} $-almost surely essentially self-adjoint on the Schwartz
space $ \mathcal{S}(\mathbb{R}^3) $ of rapidly decreasing,
arbitrarily often differentiable functions on $ \mathbb{R}^3 $.
For a wealth of information on these and related questions on random Schr\"odinger operators,
see \cite{Kir89,CaLa92,PaFi92,St01}.

As another consequence, the integrated density of states may be defined by the expectation value
\begin{equation}\label{def:N}
  N(E) := \int_{\Omega} \mathbb{P}(d\omega) \;\, \Theta\big( E - H(V_\omega) \big)(x,x).
\end{equation}
Here
$ \mathbb{R}^3 \times  \mathbb{R}^3 \ni (x, y) \mapsto \Theta\big( E - H(V_\omega) \big)(x,y) $ denotes
the continuous integral kernel (see e.g.  \cite{BHL00,BML02}) of the spectral
projection $ \Theta\big( E - H(V_\omega)\big) $ of
$ H(V_\omega) $ associated with the half-line $ ]-\infty,E [ \, \subset \mathbb{R} $. Due to magnetic
translation invariance, the r.h.s.\ of (\ref{def:N}) is independent of the chosen $ x \in \mathbb{R}^3 $.
For a background, alternative equivalent definitions and further properties of $ N $ in this and more general situations,
see \cite{Kir89,CaLa92,PaFi92,HLMW01,LMW02} and references therein.

Additionally to (\ref{Ass1}), we will suppose that $ U $ has an \emph{anisotropic} algebraic decay at infinity
\begin{equation}\label{Ass2}
    \lim_{| x | \to \infty }\, \left\|  \big(|x_\perp|^\alpha , |x_3|^\gamma \big) \right\|_{2/\beta}
    \, U\left(  x_\perp  , x_3 \right)= g \; > 0,
\end{equation}
where we used the notation
$  \left\| c \right\|_{2/\beta} := \left( | c_1 |^{2/\beta}  + |c_2|^{2/\beta}
          \right)^{\beta/2} $ ($ =\max(|c_1|,|c_2|)$ if $\beta =0$)
for the $ 2/ \beta $-pseudo-norm of a vector $ c = (c_1 , c_2 ) \in \mathbb{R}^2 $ and
$ x_\perp := (x_1 , x_2 ) \in \mathbb{R}^2 $ for the co-ordinate perpendicular to the
magnetic field. For all $\beta\ge 0$, integrability of $U$ at infinity is equivalent to
$ \alpha > 2 $  and  $ \gamma > \alpha / (\alpha -2) $.  In particular, for isotropic decay of $ U $, which corresponds to
$\alpha=\beta = \gamma$, integrability requires $\alpha>3$.
\smallskip

%\subsection{Statement of Results}
Our results on the magnetic Lifshits tails for slow anisotropic algebraic decay of $ U $ are summarized in the following
%
%
%%%%%%%%%%%%%%%%%%%%%%%%%%%%%%%%%%%%%%%%%%%%%%%%%%%%%%%%%%%%%%%%%%%%%%%%%%%%%%%%%%%%%%%%%%%%%%%%%%%%%%%%%%%%%%%
\begin{theorem}\label{Thm}
  For a positive impurity potential satisfying assumptions~(\ref{Ass1}) and (\ref{Ass2}) with some $ g > 0 $, $ \alpha > 2 $,
  $  \beta > 0 $ and
  \begin{equation}\label{Ass}
    \alpha/(\alpha -2)  < \gamma < 3 \alpha/(\alpha -2),
  \end{equation}
  the leading low-energy fall-off of
  the integrated density of states $ N $ is independent of the magnetic field strength $ B > 0 $ and reads
  \begin{equation}\label{Eq:Thm}
    \lim_{E \downarrow 0} \, E^{\frac{3}{\eta - 3}} \; \log N\!\left( \frac{B}{2} + E \right) = -
    C .
  \end{equation}
  Here we have introduced the two constants $ \eta := 3 \alpha \gamma /\left( 2\gamma + \alpha \right) $ and
  \begin{displaymath}
     C
     := \frac{\eta - 3}{3} \, g^{\frac{3}{\eta-3}}
     \left[
       2 \pi \varrho  \, \frac{\beta}{\alpha \gamma}
       \frac{\Gamma\!\left(\frac{\beta}{\alpha}\right) \,
         \Gamma\!\left(\frac{\beta}{2\gamma}\right)}{\Gamma\!\left(\frac{3\beta}{2\eta}\right)}
       \Gamma\!\left(\frac{\eta - 3}{\eta}\right)
     \right]^{\frac{\eta - 3}{\eta}} .
  \end{displaymath}
  $\mbox{}$
\end{theorem}
%%%%%%%%%%%%%%%%%%%%%%%%%%%%%%%%%%%%%%%%%%%%%%%%%%%%%%%%%%%%%%%%%%%%%%%%%%%%%%%%%%%%%%%%%%%%%%%%%%%%%%%%%%%%%%%

    For given value of $ \alpha \, ( > 2 ) $ and $ \gamma \, ( > \alpha / (\alpha - 2) )$,
    the parameter $ \beta > 0 $ fine tunes the
    degree of anisotropy of the decay of $ U $ by selecting different pseudo-norms
    in (\ref{Ass2}). Thanks to equivalence of these pseudo-norms,
    the choice of $ \beta $
    does not effect the order of the decay of $ U $
    and hence not that of $ \log N $. More precisely, $ \beta $
    does not enter the so-called
    \emph{Lifshits exponent}
    \begin{equation}
      - \lim_{E \downarrow 0} \frac{\log \left|\, \log N\!\left( \frac{B}{2} + E \right) \right| }{\log E} = \frac{3}{\eta - 3}
    = \frac{2\gamma + \alpha}{\alpha \gamma -2\gamma -\alpha},
    \end{equation}
    but only the Lifshits constant $ C $.
    In the limit $ \beta \downarrow 0 $ where
    $  \| c   \|_{2/\beta}
    \to \max \{ |c_1 | , | c_2 | \} $,
    the Lifshits constant converges to $ C \to \big( \eta/3 - 1 \big) g^{3/(\eta - 3)}
    [ 6 \pi \varrho \, \Gamma( 1 - 3/\eta ) / \eta ]^{ 1 - 3/\eta} $.
    The subsequent proof shows that Theorem~\ref{Thm} remains valid in this limiting case
    with the above value of the Lifshits constant~$ C $.

In all of the above cases, the Lifshits tails sensitively depend on the details of the decay of $ U $ and
are classical in character. Indeed,
the corresponding \emph{classical integrated density of states}
\begin{equation}\label{classIDOS}
  N_{\rm cl}(E) := \frac{\sqrt{2}}{3 \pi^2} \int_{\Omega} \mathbb{P}(d \omega) \;\,
  \big( E - V_\omega(0) \big)^\frac{1}{2}  \; \max\!\big\{  E - V_\omega(0) , 0 \big\}
\end{equation}
(cf.\ \cite[Eq.~(2.14)]{HLW99}) has the same leading low-energy tail as $ N $, that is,
$ \lim_{E \downarrow 0} \log N( B/2 + E ) / \log N_{\rm cl}(E) = 1 $.

In the extreme anisotropic limit $ \alpha \to \infty $, condition~(\ref{Ass})
turns into $ 1 < \gamma < 3 \, (= 1 + 2) $ while $ \eta \to  3 \gamma $ and
$ C \to (\gamma - 1 )\,  g^{1/(\gamma- 1) } ( 2 \pi \varrho \,
\Gamma\left( 1- 1/\gamma\right) / \gamma )^{1 - 1 / \gamma} $.
Comparing these limiting values with results in \cite{Pas77} and \cite[Cor.~9.14]{PaFi92},
the Lifshits tails (\ref{Eq:Thm}) are seen to asymptotically coincide
with the corresponding classical tails
in one space dimension for impurities with concentration $ \pi \varrho $ and algebraically
decaying $ U $ (with exponent $ \gamma $).
This is plausible from the long-distance tails of $ U $ which develop
in the direction parallel to the magnetic field
in this limit.  %\attention{Was wird aus dem Argument f\"ur $ B = 0 $?}
The quantum particle is therefore effectively confined to a one-dimensional motion.
Thus the one-dimensional picture sketched at the end of the Introduction correctly captures the strongly anisotropic case $\alpha \to \infty $,
but not the case where $ U $ has \emph{isotropic} algebraic decay,
that is, the case $ \alpha = \beta = \gamma \, (= \eta) $ for which Theorem~\ref{Thm} yields the following

\begin{corollary}\label{Cor}
Assume that $ \lim_{| x | \to \infty } | x |^\alpha \, U(x) = g > 0 $ with some $ 3 < \alpha < 5 $. Then
\begin{equation}\label{eq:istrop}
  \lim_{E \downarrow 0} \, E^{\frac{3}{\alpha - 3}} \; \log N\!\left( \frac{B}{2} + E \right)
  = - \frac{\alpha - 3}{3} \, g^{\frac{3}{\alpha-3}}  \left[
    \frac{4 \pi \varrho}{\alpha} \, \Gamma\!\left(\frac{\alpha - 3}{\alpha}\right)
  \right]^{\frac{\alpha - 3}{\alpha}}.
\end{equation}
\end{corollary}

In the isotropic case the tails (\ref{eq:istrop}) coincide for \emph{all} values of $ 3 < \alpha < 5 \, ( = d + 2 ) $
with the corresponding classical tails for $ B=0 $, cf.\ the Introduction and \cite{Pas77,PaFi92}.
This is different for anisotropic decay of $ U $.
A straightforward modification of the subsequent proof
shows that (\ref{Eq:Thm}) remains valid for $ B = 0 $ if
\begin{equation}
  \frac{\alpha}{\alpha - 2} < \gamma <
  \left\{
    \begin{array}{l@{\qquad\mbox{if}\quad}l}
      \displaystyle\frac{3\alpha}{\alpha -2} & 2 < \alpha \leq 5, \\[2ex]
      \displaystyle\frac{\alpha}{\alpha -4}  & 5 < \alpha,
    \end{array}
  \right.
\end{equation}
(see also \cite{Pas77} and \cite[Cor.~9.14]{PaFi92} for the isotropic case $ \alpha = \beta = \gamma $.)
Accordingly, the validity of (\ref{Eq:Thm}) requires $ B > 0 $ in case $ \alpha > 5 $.
This resembles the two-dimensional situation for which the authors
of \cite{BHKL95} showed that quantum effects in the Lifshits tail
are suppressed in the presence of a magnetic field.
%
%
%\Hmm{Ich denke, mann sollte auf jeden Fall herausstreichen, dass das naive Bild keine klassischen L-Tails erlaubt!!}
 % Maybe most strikingly, Corollary \ref{Cor}  is in \emph{strong contrast} to some naive,
%  but often very  helpful heuristics for quantum particles in a constant magnetic field: the motion perpendicular to the magnetic
%  field is confined and the particle can move freely only parallel to the field lines. Thus, one likes to think that the
%  effective dimension of the problem is $1$.
%  In particular, recalling the dimension dependence for the classical asymptotics for vanishing magnetic field, one is led to
%  predict that in order to have classical asymptotics,
%  the (algebraic) decay of the potential $ U $ must satisfy  $\alpha< d_{\mathrm{eff}}+2=1+2=3$.
%  However, in order that the Poissonian  potential
%  is well-defined, one needs $\alpha>3$.
%  So, arguing this way, classical asymptotics seem to be impossible for isotropic potentials. But the Corollary \ref{Cor}  shows
%  that this point of view is wrong for isotropic potentials. Note that according to Remark \ref{rem:one}.\ref{rem:extremeanisotropic},
%  the one-dimensional picture correctly captures the strongly anisotropic case, however.

\begin{remark}
  Following \cite{HLW00},
  Corollary~\ref{Cor} possesses a natural (and straightforward) extension to impurity potentials $ U $ with
  (slow) regular isotropic $ (F, \alpha) $-decay in the sense that
  there exists some $ 3 < \alpha < 5 $
  such that $ \lim_{|x| \to \infty} F\big( 1/ U(x) \big) = 1 $ for some positive function $ F $
  which is regularly varying (at infinity) with index $ 1 / \alpha $, cf.\ \cite[Def.~3.5]{HLW00}.
  Denoting by $ f^\# $ the de Bruijn conjugate \cite[p.~29]{BGT89} of the function
  $ t \mapsto f(t) := \big[ t^{-1/\alpha} F(t) \big]^{3 \alpha/(3 - \alpha)}  $,
  the corresponding Lifshits tails read
  \begin{displaymath}
    \lim_{E \downarrow 0} \, \frac{\log N\big( \frac{B}{2} + E \big)}{E^{\frac{3}{3 - \alpha}} f^\#\big(E^{\frac{3}{3 - \alpha}}\big)}
    = - \frac{\alpha - 3}{3} \, \left[
    \frac{4 \pi \varrho}{\alpha} \, \Gamma\!\left(\frac{\alpha - 3}{\alpha}\right)
  \right]^{\frac{\alpha - 3}{\alpha}}.
  \end{displaymath}
  This extension constitutes the analogue of \cite[Thm.~3.8]{HLW00} in three space dimensions.
\end{remark}

\section{Proof}
The strategy of our proof of the classical Lifshits tails in
Theorem~\ref{Thm} goes back to \cite{Lut76,Pas77} and has been adopted to the magnetic
setting in \cite{BHKL95,HLW99,HLW00}. Instead of the leading low-energy fall-off of $ N $, we will investigate
the behaviour of its (shifted) Laplace transform
\begin{displaymath}
  \widetilde N(t) := \int_0^\infty N\!\left( d E + \frac{B}{2} \right) \, e^{-t E}, \qquad t > 0,
\end{displaymath}
for long ``times'' $ t $.
De Bruijn's Tauberian theorem \cite[Thm.~4.12.9]{BGT89} (see also \cite[Thm.~9.7]{PaFi92} or \cite[App.~B]{HLW00})
ensures that (\ref{Eq:Thm}) is equivalent to
\begin{align}\label{eq:Tauber}
  \lim_{t \to \infty} t^{-\frac{3}{\eta}} \, \log \widetilde N(t)
  & =  - 2 \pi  \varrho \; g^{\frac{3}{\eta}} \frac{\beta \eta}{3 \alpha \gamma}
      \frac{\Gamma\!\left(\frac{\beta}{\alpha}\right) \, \Gamma\!\left(\frac{\beta}{2\gamma}\right)}{\Gamma\!\left(\frac{3\beta}{2\eta}\right)}
      \Gamma\!\left(\frac{\eta - 3}{\eta}\right) \\
  & =  - \varrho \, \int_{\mathbb{R}^3}\! d x \, \left( 1 -
    e^{-  u(x) }\right) . \notag
\end{align}
Here the last equality results from an elementary (but somewhat lengthy)
calculation of the last integral which is defined in terms of the function $ u: \mathbb{R}^3 \to [ 0 , \infty [ $ given by
\begin{equation} \label{eq:defu}
  u(x) := g  \left(  |x_\perp|^{\frac{2\alpha}{\beta}}  + |x_3|^{\frac{2\gamma}{\beta}}
          \right)^{- \frac{\beta}{2}} = \frac{g}{  \left\|  \big(|x_\perp|^\alpha , |x_3|^\gamma \big) \right\|_{2/\beta}}.
\end{equation}
In order to determine the long-time behaviour of $ \widetilde N $ we use the following
upper and lower bounds. They are a straightforward extension of the ones in \cite[Basic Inequalities~3.1]{BHKL95}
for two to three space dimensions, see also \cite[Prop.~5.3]{War01}.
\begin{proposition}
Let $ \psi \in \mathcal{S}(\mathbb{R}^3) $ be normalized according to the standard scalar product
$ \langle \psi , \psi \rangle = 1 $ on $ {\rm L^2}(\mathbb{R}^3) $. Moreover assume that $ \psi $ is real-valued
and centred in the sense that
$ \int_{\mathbb{R}^3} d x \, |\psi(x) |^2 \, x = 0 $.
Then the sandwiching estimate
\begin{multline}\label{eq:bounds}
    \frac{1}{\sqrt{(2 \pi t)^{3}}} \, \exp\left[ - t\,  \big\langle \psi  , H(0) \, \psi \big\rangle
      - \varrho \int_{\mathbb{R}^3} \! d x \, \left( 1 - e^{-t \int_{\mathbb{R}^3}  d y \, | \psi (x - y) |^2 \, U(y) } \right) \right] \\
      \leq e^{-t B/2} \, \widetilde N(t)
      \leq \frac{B}{4 \pi \sqrt{2 \pi t}\, \sinh(t B/2)} \, \exp\left[  -
        \varrho \int_{\mathbb{R}^3} \! d x \, \left( 1 - e^{-t U(x)} \right) \right]
\end{multline}
holds for all values of the magnetic field strength $ B \geq 0 $.
\end{proposition}
In the next two subsections we will show that,
after choosing the variational state-vector $ \psi $ properly, the bounds (\ref{eq:bounds})
asymptotically coincide in the situation of Theorem~\ref{Thm}.
This will complete our proof of (\ref{eq:Tauber}) and hence of Theorem~\ref{Thm}.
Note that the exponential factor in the upper bound coincides up to a factor of $ \sqrt{(2 \pi t)^3} $
with the (unshifted) Laplace transform of $ N_{\rm cl} $.
Therefore, (\ref{eq:Tauber}), and hence (\ref{Eq:Thm}), is indeed a classical asymptotics.
\subsection{Asymptotic evaluation of the upper bound}
The upper bound in (\ref{eq:bounds}) implies
\begin{displaymath}
  \limsup_{t \to \infty} \, t^{-\frac{3}{\eta}}  \, \log  \widetilde N(t)
  \leq
  - \varrho \, \liminf_{t \to \infty} \, t^{-\frac{3}{\eta}} \int_{\mathbb{R}^3}\! d x \, \left( 1 - e^{-t U(x)}\right).
\end{displaymath}
Using the substitution $ (x_\perp , x_3 ) \mapsto \big( t^{\frac{1}{\alpha}} x_\perp ,  t^{\frac{1}{\gamma}} x_3 \big) $
and subsequently employing Fatou's lemma together with the pointwise convergence
$ \lim_{t \to \infty } t  \, U\big( t^{\frac{1}{\alpha}} x_\perp ,  t^{\frac{1}{\gamma}} x_3 \big) = u(x) $
valid for all $ x \neq 0 $, we thus conclude that
\begin{equation}\label{eq:asyup}
  \limsup_{t \to \infty} \, t^{-\frac{3}{\eta}}  \, \log \widetilde N(t)
  \leq - \varrho  \, \int_{\mathbb{R}^3}\! d x \, \left( 1 -
    e^{-  u(x) }\right).
\end{equation}

\subsection{Asymptotic evaluation of the lower bound}
We choose the variational state-vector in our lower bound (\ref{eq:bounds})
as follows
\begin{equation}\label{eq:defpsi}
  \psi(x) := \sqrt{\frac{B}{2\pi}} \exp\left(- \frac{B}{4} \, |x_\perp |^2 \right)
  \, \frac{1}{\sqrt{t^{{\sigma}}}}
  \, \varphi\left( \frac{x_3}{t^{\sigma }}  \right), \qquad t > 0.
\end{equation}
It is the time-dependent product of the centred Gaussian
in the lowest Landau-level eigenspace and some real-valued, centred,
arbitrarily often differentiable,
compactly supported $ \varphi \in \mathcal{C}_c^\infty(\mathbb{R}) $ scaled by $ t^\sigma $.
We will pick $ \sigma > 0 $ later at our convenience. To ensure normalization of $ \psi $ we further
assume $ \varphi $ to be normalized with respect the the standard scalar product $ \langle \varphi , \varphi \rangle = 1 $
on $ {\rm L}^2(\mathbb{R}) $.
Accordingly, the first term in the exponent on the l.h.s.\ in (\ref{eq:bounds}) equals
\begin{displaymath}
  t  \; \big\langle \psi  , H(0) \, \psi \big\rangle = \frac{t B}{2} + t^{1- 2 \sigma}
  \, \big\langle \varphi  , H_3(0) \, \varphi \big\rangle
\end{displaymath}
where $ H_3(0) := - \frac{1}{2} \frac{\partial^2}{\partial x_3^2} $.
Using the substitution $ (x_\perp , x_3 ) \mapsto \big( t^{\frac{1}{\alpha}} x_\perp ,  t^{\frac{1}{\gamma}} x_3 \big) $,
the second term in the exponent on the l.h.s.\ in (\ref{eq:bounds}) may be expressed in terms of
the one-parameter family $ \left\{ \delta_t \right\}_{t > 0 } $ of probability densities on $ \mathbb{R}^3 $
given by
\begin{equation}\label{def:d}
  \delta_t\left(x\right) := t^{\frac{2}{\alpha}+\frac{1}{\gamma}} \left| \psi\left( t^{\frac{1}{\alpha}} x_\perp , t^{\frac{1}{\gamma}} x_3 \right) \right|^2.
\end{equation}
We thus arrive at
\begin{multline}\label{eq:lowerasy}
  \liminf_{t \to \infty} \, t^{-\frac{3}{\eta}}  \, \log \widetilde N(t)
  \geq   - \limsup_{t \to \infty} \;  t^{1- 2 \sigma - \frac{2}{\alpha} - \frac{1}{\gamma}} \,
  \big\langle \varphi  , H_3(0) \, \varphi \big\rangle \\
  - \varrho \,  \limsup_{t \to \infty}\;  \int_{\mathbb{R}^3} \! d x \,
  \left[ 1 - \exp\left(-t  \int_{\mathbb{R}^3} \! d y \, \delta_t\left(x- y \right) \, U\left(t^\frac{1}{\alpha} y_\perp, t^\frac{1}{\gamma} y_3 \right)    \right)\right] .
\end{multline}
Since $ 0< 1 - \frac{2}{\alpha} - \frac{1}{\gamma} < \frac{2}{\gamma} $ by assumption, we may pick $ \sigma > 0 $ such that
\begin{displaymath}
  1 - \frac{2}{\alpha} - \frac{1}{\gamma} < 2 \sigma < \frac{2}{\gamma} .
\end{displaymath}
Therefore the first term in the r.h.s.\ of (\ref{eq:lowerasy}) vanishes. Thanks to Lemma~\ref{Lemma} below
the second term my be handled with the help of the dominated convergence theorem. Altogether, we thus conclude
\begin{displaymath}
  \liminf_{t \to \infty} \, t^{-\frac{3}{\eta}}  \, \log \widetilde N(t)
  \geq - \varrho  \, \int_{\mathbb{R}^3}\! d x \, \left( 1 -
    e^{-  u(x) }\right) ,
\end{displaymath}
which, together with (\ref{eq:asyup}), completes the proof of (\ref{eq:Tauber}).

\subsection{Auxiliary lemma}
\begin{lemma}\label{Lemma}
  Let $ 0 \leq \sigma < 1/\gamma $. Then
  \begin{equation}\label{eq:lemma}
    \limsup_{t \to \infty} \, t \int_{\mathbb{R}^3}\! d y \;\delta_t\left(x - y\right)\, U\left( t^{\frac{1}{\alpha}} y_\perp ,  t^{\frac{1}{\gamma}} y_3 \right)
     \leq u(x)
  \end{equation}
  for all $ x \neq 0 $.
\end{lemma}
\begin{proof}
  We pick $ 0 < \varepsilon < 1 $ and split the convolution in the l.h.s.\ of (\ref{eq:lemma})
  into two integrals with domains of
  integration inside and outside the ball $ B_{\varepsilon |x|}^{\, (x)} $ centred at $ x $ with radius $  \varepsilon | x | $.
  Using the fact that the $ \delta_t $ is a probability density, the first part is estimated as follows
  \begin{equation}\label{eq:lemma1}
    t \int_{B_{\varepsilon |x|}^{\, (x)}}\!\! d y \;
    \delta_t\left(x - y\right)\, U\left( t^{\frac{1}{\alpha}} y_\perp ,  t^{\frac{1}{\gamma}} y_3 \right)
    \leq
    \esssup_{y \in B_{\varepsilon |x|}^{\, (x)}} \, t \,  U\left( t^{\frac{1}{\alpha}} y_\perp ,  t^{\frac{1}{\gamma}} y_3 \right) .
  \end{equation}
  Since $ | y | \geq (1 - \varepsilon) | x | > 0 $, we may further estimate the r.h.s.\ with the help of
  the inequality
  \begin{equation}
    t \,  U\left( t^{\frac{1}{\alpha}} y_\perp ,  t^{\frac{1}{\gamma}} y_3 \right)
    \leq
    (1 - \varepsilon) \, t \, u\left( t^{\frac{1}{\alpha}} y_\perp ,  t^{\frac{1}{\gamma}} y_3 \right)
    = (1 - \varepsilon) \, u(y) ,
  \end{equation}
  valid for sufficiently large $ t > 0 $ by assumption \eqref{Ass2} on $ U $
  and the definition \eqref{eq:defu} for $u$.
  To estimate the second part we employ the inequality
  \begin{align}
   t \int_{\mathbb{R}^3 \setminus B_{\varepsilon |x|}^{\, (x)}}\!\!\! d y \,
    \;  \delta_t\left(x - y\right) \, U\left( t^{\frac{1}{\alpha}} y_\perp ,  t^{\frac{1}{\gamma}} y_3 \right)
     & \leq t  \! \sup_{y \notin B_{\varepsilon |x|}^{\, (x)}} \!\! \delta_t\left(x - y\right) \,
    \int_{\mathbb{R}^3}\! d z \, U\left( t^{\frac{1}{\alpha}} z_\perp ,  t^{\frac{1}{\gamma}} z_3 \right)  \notag \\
    & =  t^{1-\frac{3}{\eta}}  \sup_{y \notin B_{\varepsilon |x|}^{\, (0)} } \!\!  \delta_t\left(y\right) \,
   \int_{\mathbb{R}^3}\! d z \, U\left(z \right) .
  \end{align}
  The upper limit of the r.h.s.\ vanishes since
  \begin{equation}\label{eq:supweg}
    \limsup_{t \to \infty} \,  t^{1-\frac{3}{\eta}} \!\! \sup_{y \notin B_{\varepsilon |x|}^{\, (0)} } \!\!\! \delta_t\left(y\right)
    = 0.
  \end{equation}
  For a proof of this assertion we distinguish two cases.
  In the first case, where $ | y_3 | \geq \varepsilon | x | / \sqrt{2} \, ( > 0 ) $,
  we have  $ \varphi\big( t^{\frac{1}{\gamma}-\sigma} y_3  \big) = 0 $ and hence $ \delta_t(y) = 0 $ for sufficiently large $ t $,
  since $ \supp \varphi $ is compact and $  t^{\frac{1}{\gamma} - \sigma} $ grows unboundedly.
  In the second case, where $ y \notin B_{\varepsilon |x|}^{\, (0)} $ and
  $ | y_3 | < \varepsilon | x | / \sqrt{2} $ such that $ | y_\perp | \geq \varepsilon | x | / \sqrt{2} $, the supremum in
  (\ref{eq:supweg}) decreases exponentially fast in time $ t $
  thanks to the Gaussian decay of $ | \psi(y) |^2 $ in the $ y_\perp $-direction, cf.\ (\ref{eq:defpsi}).
  Altogether (\ref{eq:lemma1})--(\ref{eq:supweg}) implies
  \begin{equation}
     \limsup_{t \to \infty} \, t \int_{\mathbb{R}^3}\! d y \,   \delta_t\left(x - y\right)
 U\left( t^{\frac{1}{\alpha}} y_\perp ,  t^{\frac{1}{\gamma}} y_3 \right)
     \leq (1 - \varepsilon) \,  \sup_{y \in B_{\varepsilon |x|}^{\, (x)}} \, u(y).
  \end{equation}
  Taking the limit $ \varepsilon \downarrow 0 $, the r.h.s.\ converges to $ u(x) $ for all $ x \neq 0 $
  since $ u $ is continuous on $\mathbb{R}^3\setminus \{0\}$.
\end{proof}

\section*{Acknowledgement}
It is a pleasure to thank Hajo Leschke and Georgi Raikov for helpful
and stimulating discussions. This work was partially supported by the SFB
237 \emph{Unordnung und grosse Fluktuationen}.

\end{document}